\documentclass[10pt,english,amsmath,amssymb,amsfonts,floatfix,aps,superscriptaddress,twocolumn,pra]{revtex4-2}
\usepackage{amsmath,amssymb,amsfonts,graphicx,physics,bm,mathtools,microtype,xcolor,amsthm,booktabs}
\usepackage[
  colorlinks=true,
  breaklinks=true,
  hypertexnames=false,
  urlcolor=blue,
  citecolor=blue,
  linkcolor=blue
]{hyperref}
\usepackage{enumitem}
\setlist{nosep}

\binoppenalty=\maxdimen
\relpenalty=\maxdimen

\newcommand{\Dicke}[1]{#1}
\newcommand{\Dickemat}{\bm M}
\newcommand{\Momentmat}{\overline{\bm M}}
\newcommand{\me}{\mathrm{e}}
\newcommand{\iu}{\mathrm{i}}
\newcommand{\ind}{k}
\newcommand{\Dickeket}{\ket{\Dicke{\ind{}}}}
\newcommand{\Ei}{\varepsilon_i}
\newcommand{\deltat}{\delta}

\newcommand{\appref}{App.~}
\newcommand{\li}{\ell}
\newcommand{\indexset}[2]{#1_{0},\dots,#1_{#2}}

\newcommand{\iind}{\indexset{i}{r-1}}
\newcommand{\lind}{\indexset{\li}{r-1}}

\newcommand{\orderedsum}[1]{\sum_{\substack{0\leq #1_0\leq #1_1\leq \dots\leq #1_{r-1} \\ \sum #1_j=r(r-1)/2}}}

\DeclarePairedDelimiter\ceil{\lceil}{\rceil}
\DeclarePairedDelimiter\floor{\lfloor}{\rfloor}

\DeclareMathOperator{\sgn}{sign}

\begin{document}

\title{Absence of Entanglement Growth in Dicke Superradiance}
\date{\today}
\author{N.~S.~Bassler}
\affiliation{Max Planck Institute for the Science of Light, Staudtstra\ss e 2, D-91058 Erlangen, Germany}
\affiliation{Department of Physics, Friedrich-Alexander-Universit\"at Erlangen-N\"urnberg, Staudtstra\ss e 7, D-91058 Erlangen, Germany}
\affiliation{TU Darmstadt, Institute for Applied Physics, Hochschulstra\ss e 4A, D-64289 Darmstadt, Germany}

\begin{abstract}
Dicke superradiance describes an ensemble of $N$ permutationally invariant two-level systems collectively emitting radiation with a peak radiated intensity scaling as $N^2$. 
Individual Dicke states are typically entangled. However, the density matrix during superradiant decay is a mixture of such states, raising the subtle question of whether the total state is entangled or separable.
We resolve this by showing analytically that for any $N$, starting from the fully excited state, the collective decay preserves separability for all times. 
This answers a longstanding question on the role of entanglement in Dicke superradiance and underscores that, despite collective dissipation, separable states remain separable under these dynamics.
\end{abstract}

\maketitle

\section{Introduction} 

For seven decades a basic question has shadowed \emph{Dicke superradiance}\,\cite{dickeCoherenceSpontaneousRadiation1954a}:  
Can \emph{purely collective} spontaneous emission ever generate genuine multipartite entanglement?  
The answer matters because controlled superradiant decay is now a working tool in state‑of‑the‑art platforms—from bad‑cavity optical clocks\,\cite{Bohr2024}, through Rydberg “superatoms” \cite{Stiesdal2020}, to superconducting transmon pairs in 3D cavities\,\cite{Mlynek2014}.  
A rigorous verdict therefore sets an experimental baseline: any entanglement detected during undriven decay must stem from non‑collective ingredients such as dipole–dipole shifts, inhomogeneous couplings, or external driving.

Dicke states---the permutationally symmetric eigenstates of the total spin operators $S^2$ and $S_z$---generically exhibit multipartite entanglement \cite{toth2007,duan2011,lucke2014} except for the fully excited and ground state. Thus, Dicke superradiance is often associated with the generation of entanglement \cite{somech2023,wolfe2014arxiv}.

Because the system becomes a mixture of entangled states during decay, it remains unclear whether the overall state is entangled. There is indeed growing evidence that symmetric Dicke dynamics does not generate entanglement but destroys åit. The challenge lies in the fact that deciding whether a mixed quantum state is separable is NP-hard in general \cite{gurvits2003,korbicz2005,korbicz2006,yu2021}, and entanglement detection for mixed, multipartite states lacks a simple universal criterion \cite{horodecki1996,horodecki1998,peres1996,acin2001}.

Wolfe and Yelin verified numerically (up to $N=10$) that the partial‑transpose (PPT) test remains satisfied throughout the decay and conjectured that this implies separability for all $N$ \cite{wolfe2014}. Yu \cite{yu2016} later proved analytically that any mixture of Dicke states is separable \emph{iff} it is PPT, thereby characterizing the \emph{static} diagonal‑symmetric set for arbitrary $N$. Yu’s proof, however, does not address whether the Lindblad superradiant trajectory stays inside that separable set. Hence, a fully analytic, all‑$N$ answer to the \emph{dynamical} question has remained open.

The problem is greatly simplified since the states evolving under Dicke dynamics are permutationally invariant and diagonal (i.e., no single-particle coherence between the excited and ground state) at all times, allowing for analytical solutions of the population dynamics \cite{Lee1977a,Lee1977b,Holzinger2024,Holzinger2025}. This structure allows the use of specialized separability criteria tailored to such states \cite{eckert2002,toth2009,toth2010,augusiak2012,yu2016,quesada2017,rutkowski2019}.

Understanding how entanglement is generated, preserved, or destroyed in open quantum systems remains a fundamental challenge in quantum information science. Despite significant efforts devoted to entanglement generation in driven or coherent settings \cite{guehne2009,amico2008}, the subtleties of entanglement dynamics under purely dissipative collective processes, such as Dicke superradiance, have remained largely unexplored \cite{krauter2011,braun2002,verstraete2009}. Our work directly addresses this knowledge gap, providing analytic clarity and resolving whether the simplest form of collective dissipation alone can generate entanglement. It is therefore highly relevant for both fundamental quantum theory and practical quantum information applications.

In this work, we use techniques from real analysis to address separability. Specifically, we apply an exact mapping to the truncated moment problem \cite{schmudgen2017moment}. Prior work also relates separability criteria of symmetric states to the moment problem \cite{quesada2017,rutkowski2019,yu2016}.
The CP‑matrix approach of Tura et al. \cite{tura2018} gives an equivalent static criterion for diagonal‑symmetric states. We extend the analysis to the full time evolution of Dicke superradiance.

We show that the population dynamics of Dicke superradiance corresponds to the evolution of a set of moments, and that if the initial state is separable, the dynamics preserves the structure required for a classical moment representation. Considering the time evolution in the lowest nonvanishing order in time enables a full analytic proof that Dicke superradiance does not generate entanglement from initially separable states. 

In the Dicke limit the system state never acquires off‑diagonal components in the Dicke basis, so all dynamics are confined to the $(N+1)$‑dimensional symmetric manifold.  This explains why \emph{symmetric‑subspace simulations} (costing only $\mathcal O(N^2)$ or better) capture the exact evolution.  Moreover, as shown in the companion Letter\,\cite{Schachenmayer2025}, one can unravel the master equation into at most $\lceil (N+1)/2\rceil$ separable spin‑coherent trajectories, yielding dramatic further speedups without loss of accuracy.  Finally, by turning off any coherent drives, experiments in bad‑cavity clocks, Rydberg arrays, and transmon ensembles can use the absence of entanglement as a \emph{calibration baseline}; re‑introducing a drive then cleanly isolates genuine many‑body correlations.

\section{The problem statement} 

We consider a system of $N$ identical two-level systems undergoing collective spontaneous emission in the Dicke limit. The dynamics is governed by the master equation
\begin{equation}\label{eq:lindblad_time}
    \dot{\rho} = \Gamma\left[S \rho S^\dagger - \frac{1}{2}\left(S^\dagger S\rho+\rho S^\dagger S\right)\right],
\end{equation}
where we set the collective decay rate $\Gamma=1$, and $S = \sum_{k=1}^N \sigma_k$ is the collective lowering operator with the single-particle lowering operator $\sigma_j=\outerproduct{g}{e}_j$ taking the two-level system $j$ from the excited to the ground state. Starting in the fully excited state $\ket{eee\dots e}$, the system remains confined to the symmetric subspace of maximal total spin $S=N/2$, spanned by the Dicke states $\Dickeket{}$ with $\ind{}=0,1,\dots,N$ defined by $S^\dagger S\Dickeket{}=h_{\ind}\Dickeket{}$ where $h_\ind{} = \ind{}(N-\ind{}+1)$. Writing the density matrix as a mixture of Dicke states $\rho(t)=\sum_{k=0}^Np_\ind{}(t)\outerproduct{\ind{}}{\ind{}}$ with populations $p_k(t)$ and without coherences in the Dicke basis, the dynamics are then described by a set of coupled rate equations
\begin{equation}\label{eq:rate_equations}
    \dot{p}_\ind{}(t) = h_{\ind{}+1}\, p_{\ind{}+1}(t) - h_\ind{}\, p_\ind{}(t).
\end{equation}
A key simplification is that we consider density matrices without coherence in the Dicke basis, i.e., that the density matrix is diagonal \footnote{Starting from the fully excited state and evolving under collective decay, the state remains diagonal in the Dicke basis at all times. This structure allows for a global phase average over spin coherent states, reducing degrees of freedom in the decomposition to a single real parameter, the excitation probability $\varepsilon$. As a result, the separability problem reduces to a one-dimensional truncated moment problem.}. Analytical approaches to the above equations have shown a simple solution for the time dynamics of the population expressed as a contour integral in the complex plane \cite{Holzinger2024,Holzinger2025}.\\
We now address the central question:
\emph{Does Dicke superradiance generate entanglement from an initially separable state?}

A mixed state of \(N\) emitters is separable if and only if it can be written as a convex sum of $M$ tensor products of individual particle density operators \cite{horodecki2009}
\begin{equation}\label{eq:unravel}
    \rho(t) = \sum_{i=1}^M w_i(t)\, \rho^{(i)}_1(t) \otimes \cdots \otimes \rho^{(i)}_N(t),
\end{equation}
with \( w_i(t) \geq 0 \), \( \sum_i w_i(t) = 1 \), and \( \rho^{(i)}_j(t) \) are the single-emitter density matrices. For permutationally invariant states, the single-particle density matrices \( \rho^{(i)}_j(t) \) in each decomposition term must be identical, making each component a symmetric product state. These are precisely spin coherent states (see also Ref.~\cite{Schachenmayer2025}), of the form
\begin{equation}
\begin{aligned}
\rho^{(i)}_j(t)&=\rho(\Ei{}(t))=\outerproduct{\psi_j(\Ei)}{\psi_j(\Ei)}\\
\ket{\psi_j(\Ei)} &= \sqrt{1-\Ei}\,\ket{g}_j+\sqrt{\Ei}\,\ket{e}_j.
\end{aligned}
\end{equation}
The single-emitter density operator corresponds to a pure state with excitation probability $\Ei\in[0,1]$.

Expressing the state in the Dicke basis and relating it to the time evolution obtained from Eq.~\eqref{eq:lindblad_time}, we find that separability is equivalent to the condition
\begin{equation}\label{eq:separability_criterion_intro}
    p_\ind{}(t) = \sum_{i=1}^M w_i(t)b_{N,\ind{}}(\Ei(t)),
\end{equation}
i.e., the population vector $\bm p(t)=(p_0(t),\dots,p_N(t))^T$ is a convex combination of Bernstein basis polynomials 
\begin{equation}
    b_{N,\ind{}}(x)=\binom{N}{\ind{}} x^\ind{} (1 - x)^{N-\ind{}}
\end{equation} 
of the excitation probabilities $\Ei(t)$. Since global phase averaging removes coherences in the Dicke basis, we focus on diagonal density matrices throughout, see \appref{}\ref{app:separability}. As we already know the analytical solution for the population dynamics, we have to invert the equation above, which means that \emph{separability is thus equivalent to finding the time-evolving coefficients $w_i(t)$ and $\Ei{}(t)$ that satisfy Eq.~\eqref{eq:separability_criterion_intro}.}

\section{Proof for absence of entanglement generation} 

Before proceeding to analyze separability preservation, we recall \cite{Holzinger2024,Holzinger2025,Lee1977a,Lee1977b} from Eq.~\eqref{eq:rate_equations} that the Dicke dynamics form a simple rate equation system governed by
\begin{equation}
    \bm p(t) = \exp(\Dickemat t)\, \bm p(0),
\end{equation}
 where \( \Dickemat \) is the upper triangular rate matrix
\begin{equation}
    \Dickemat = \begin{pmatrix}
        0        & h_1     &              &  \\
        0        & -h_1    & \ddots      & \\
                 &         & \ddots      & h_N \\
                 &         & 0            & -h_N
    \end{pmatrix}.
\end{equation}

To build intuition, we first study bipartite entanglement between two atoms in the ensemble. This quantity - the two-particle negativity - is tractable analytically and provides a lower bound on any entanglement present in the system. Surprisingly, we find that even when starting from an entangled Dicke state, this quantity always decreases during the collective evolution, see \appref{}\ref{app::negativity}. Numerically, this negativity decreases monotonically on a characteristic timescale $\deltat\sim 1/(\Gamma N)$. In particular, for an initially fully excited state, the negativity always remains zero. While this analytical result and the accompanying numerical results seem to suggest that no entanglement is created during the dynamics, the negativity of the density matrix with respect to bipartitions only captures bipartite entanglement. Thus, one cannot rule out the generation of more complex multipartite entanglement. To address this, we now define a more general separability criterion. The reformulation in terms of a moment problem bypasses the need for computationally intensive optimization over all possible entangled states, instead reducing the problem to verifying the positive semidefiniteness of two matrices, making analytics tractable.

Below we give the essential argument; algebraic details, Hankel determinants, and an introduction to the truncated Hausdorff moment problem are given in \appref{}\ref{app:hausdorff}.

\section{Why collective decay cannot create entanglement}

Collective emission conserves full permutation symmetry. That symmetry forces the density matrix to stay \emph{diagonal} in the Dicke basis, i.e.\ a classical probability vector over Dicke levels. Entanglement can only appear if those probabilities \emph{cannot} be written as a mixture of spin-coherent product states. We show they always can.

\begin{figure}[t]
    \centering
    \includegraphics[width=\linewidth]{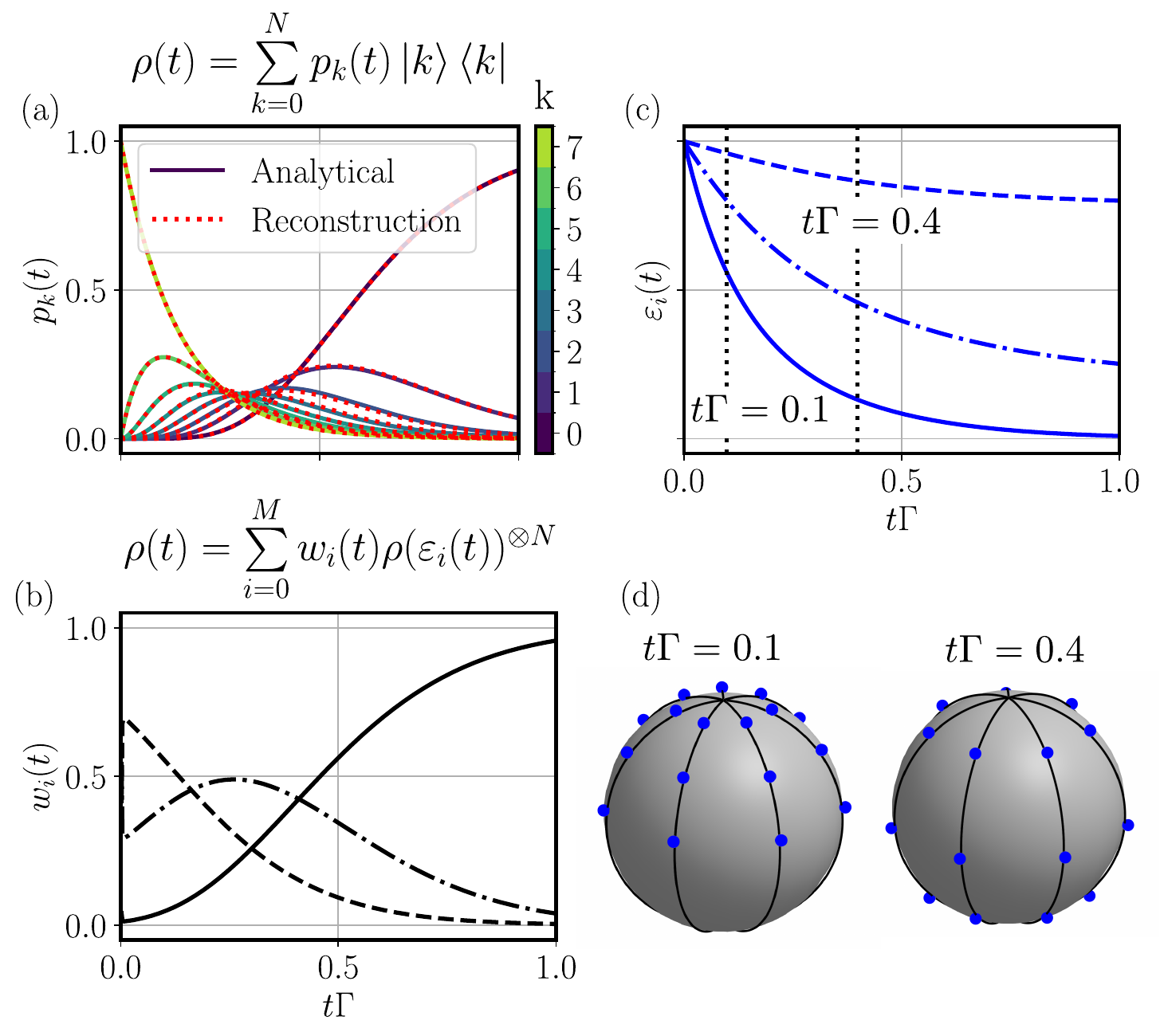}
    \caption{Coherent-state decomposition of Dicke superradiance dynamics for $N = 7$.  
    (a) Time evolution of Dicke state populations \( p_k(t) \), shown as solid lines with color indicating excitation number $k = 0,1, \dots, 7$. Red dotted lines show reconstructed populations from the moment decomposition Eq.~\eqref{eq:separability_criterion_intro}.  
    (b) Time-dependent decomposition weights $w_i(t)$ for the representation $\rho(t) = \sum_i w_i(t) \rho(\Ei(t))^{\otimes N}$, plotted in black with distinct linestyles.  
    (c) Corresponding excitation probabilities $\Ei(t)$, shown in blue with matching linestyles. Vertical dotted lines indicate the time slices used in (d).  
    (d) Bloch sphere visualization of reconstructed spin-coherent components at selected times. Each blue dot corresponds to a pure product state $\ket{\psi(\Ei, \phi)}^{\otimes N}$, where the azimuthal angle encodes the phase $\phi$ used for phase averaging. The polar angle gives the excitation probability, which decreases over time as seen in the distribution of the blue dots at $t=0.1\Gamma$ compared to $t=0.4\Gamma$.
    The full reconstruction algorithm is described in \appref{}\ref{app:moment_reconstruction}. Phase averaging is implemented explicitly via a discrete set of phases as discussed in the \appref{}\ref{app:separability}, ensuring compatibility with the diagonal form of the Dicke density matrix.}
    \label{fig:reconstructions}
\end{figure}

\begin{enumerate}
\item \emph{Diagonal trajectory}.  
   Eq.~\eqref{eq:lindblad_time} evolves the Dicke populations linearly: $\bm p(t)=e^{\bm M t}\bm p(0)$.
\item \emph{Moments instead of populations}.  
   Via the Bernstein–monomial transform $\bm m=\bm B\bm p$ we regard $m_k(t)$ as the first $N$ moments of some (unknown) distribution of the excitation probability on $x\!\in\![0,1]$.
\item \emph{Truncated Hausdorff moment problem}.  
   A distribution exists \emph{iff} two Hankel matrices $H(\bm m)$ and $\bar H(\bm m)$ remain positive-semidefinite. This is equivalent to saying “there is a spin-coherent decomposition.” Such a reconstruction is shown in Fig.~\ref{fig:reconstructions}.
\item \emph{Time-step stability}.  
   Expanding $H(\bm m)$ in a small step $\delta t$ shows that all leading minors change only in first nonzero order $\delta t^{\,r(r-1)/2}$ with a \emph{positive} prefactor (\appref{}\ref{app:all_minors}). Hence $H$ and $\bar H$ stay positive for any $\delta t\!\le\!\epsilon$ for some $\epsilon>0$ that does not depend on $x$.
\item \emph{Full trajectory}.  
   Composing a sufficient number of such steps covers all $t\!\ge\!0$, so the state is separable at every instant.
\end{enumerate}

Each blue dot in panel (d) is a spin-coherent product state; taken together, the cloud represents a classical probability distribution of Bloch vectors. Collective decay simply reshuffles probability mass toward the south pole without creating quantum correlations. The average Bloch vector reproduces the exact Dicke dynamics.

\section{Radiated intensity from the separable mixture}

Because every pure component $\rho(\varepsilon)^{\otimes N}$ in the
decomposition carries an inversion $\varepsilon$,
the instantaneous superradiant intensity can be written in closed‐form
without invoking entanglement:
\begin{equation}
I(t)=\Gamma\expval{\hat S^{\dagger}\hat S}
=\Gamma\int_0^1 f_t(\varepsilon)\bigl[N\varepsilon
+N(N-1)\,\varepsilon(1-\varepsilon)\bigr]\,d\varepsilon\,.
\end{equation}
The familiar $N^2$ scaling thus arises purely from the \emph{classical} statistics encoded in $f_t(\varepsilon)$; no quantum
correlations between emitters are required.

\section{Connection to quantum trajectories}

Eq.~\eqref{eq:unravel} may also be viewed as an \emph{unraveling}: the initial photon emission stochastically selects a spin-coherent branch, with branch probabilities $w_i(t)$.  Along every individual trajectory, the state remains \textit{fully separable}; all quantum randomness is encoded in the classical distribution $f_t(\varepsilon)$. A complementary Letter (Rosario \emph{et al.} \cite{Schachenmayer2025}) exploits this fact to build low-entanglement simulation algorithms.

\section{Experimental consequence}

Our proof assumes an initially \emph{diagonal} Dicke state and no external drive.  This is precisely the protocol used in pulsed superradiant bursts of (i) bad‑cavity optical‑clock atoms \cite{Bohr2024}, (ii) transmon pairs \cite{Mlynek2014}, and (iii) Rydberg or solid‑state emitter arrays \cite{Vasilev2024, Masson2022, Stiesdal2020}.  In such ``free‑decay'' runs, the trajectory is rigorously separable, so \emph{any} entanglement measured must originate from \emph{non‑collective} channels—dipole‑dipole shifts, spatial inhomogeneity, or dephasing.
By contrast, when the same platforms are operated in
continuous‑wave or Ramsey mode, a coherent drive maintains spin superpositions \cite{zhang2018,somech2023}; those off‑diagonal terms lie outside the Dicke‑limit Lindblad considered here and can indeed generate entanglement. Our result therefore serves as a calibration
baseline: turning off the drive should eliminate entanglement, while turning it back on isolates the genuine coherent contributions.

\section{Discussion and Conclusion} 

We have analytically shown that Dicke superradiance does not generate entanglement when starting from a separable, symmetric, and diagonal state. This settles a longstanding question about entanglement in Dicke superradiance: Collective dissipation in the Dicke limit preserves separability.

Our proof uses the structure of symmetric diagonal Dicke density matrices to map the separability question onto the truncated moment problem on $[0,1]$. By analyzing the time evolution of a product state under a small time step, we found that the resulting moments satisfy the truncated moment problem conditions. Thus, this small time evolution leaves separable states separable. Composing such steps shows that the entire trajectory remains separable.

Additionally, we introduced an entanglement measure, the sum over the absolute values of the negative eigenvalues of the Hankel matrices, in \appref{}\ref{subsec:entanglement_measure}; it vanishes iff the state is separable, decays exponentially $\propto 1/N$ in our numerics, and remains zero when starting from the fully excited state --- confirming the analytic proof.

Our analysis provides a framework for studying entanglement dynamics in collective spin models. While we have shown that separability is preserved under collective decay in the Dicke limit, entanglement can emerge in more general settings---such as emitter arrays with spatially structured decay, or in systems strongly coupled to a lossy cavity. These scenarios may feature entanglement generation or even transient entanglement transitions. Extending our approach to states with coherence, to symmetry-breaking dynamics, or models like Tavis-Cummings is a promising direction for future work. It also raises the question of whether the absence or emergence of entanglement could be confirmed directly in experiment.\\

\noindent\emph{Acknowledgments---} I thank Claudiu Genes for help with the structure of this manuscript. I thank Julian Lyne, Kai P.~Schmidt, Andreas Schellenberger, Julius Gohsrich, Lea Lenke, and Manuel Bojer for carefully proofreading the manuscript. I also thank Johannes Schachenmayer and Pedro Rosario for discussions and for the idea that sparked this project. I acknowledge financial support from the Max Planck Society and the Deutsche Forschungsgemeinschaft (DFG, German Research Foundation) -- Project-ID 429529648 -- TRR 306 \mbox{QuCoLiMa}
(``Quantum Cooperativity of Light and Matter'').

\appendix

\section{Separability Criterion and Convex Decomposition}\label{app:separability}

Consider a single-emitter pure state
\begin{equation}
\ket{\psi(\varepsilon,\phi)} = \sqrt{1-\varepsilon}\,\ket{g} + \me^{\iu\phi}\sqrt{\varepsilon}\,\ket{e},
\end{equation}
with $0\le \varepsilon\le 1$ and $\phi\in[0,2\pi)$. Its density matrix is
\begin{equation}
\begin{aligned}
    \rho(\varepsilon,\phi) &= \ket{\psi(\varepsilon,\phi)}\bra{\psi(\varepsilon,\phi)}\\
&=\begin{pmatrix}
\varepsilon & \me^{-\iu\phi}\sqrt{\varepsilon(1-\varepsilon)}\\
\me^{\iu\phi}\sqrt{\varepsilon(1-\varepsilon)} & 1-\varepsilon
\end{pmatrix}
\end{aligned}
\end{equation}
in the basis $\{\ket{e},\ket{g}\}$.

Now, take the tensor product of $N$ identical emitters
\begin{equation}
\ket{\psi(\varepsilon,\phi)}^{\otimes N} = \left(\sqrt{1-\varepsilon}\,\ket{g} + e^{\iu\phi}\sqrt{\varepsilon}\,\ket{e}\right)^{\otimes N}.
\end{equation}

Expanding this vector in the Dicke basis, we find
\begin{equation}
\ket{\psi(\varepsilon,\phi)}^{\otimes N} 
=\sum_{\ind{}=0}^{N} \sqrt{\binom{N}{\ind{}}}\, \me^{\iu\phi\ind{}}\, \varepsilon^{\ind{}/2}\,(1-\varepsilon)^{(N-\ind{})/2}\,\ket{n},
\end{equation}
leading to a density matrix
\begin{equation}
\begin{aligned}
     \rho(\varepsilon,\phi)^{\otimes N} =&{} \sum_{\ind{},\ind{}'=0}^{N} \sqrt{\binom{N}{\ind{}}\binom{N}{\ind{}'}}\, \me^{\iu\phi\,(\ind{}-\ind{}')}\, \\
     &\times\varepsilon^{\frac{\ind{}+\ind{}'}{2}}\,(1-\varepsilon)^{\frac{2N-(\ind{}+\ind{}')}{2}}\,\outerproduct{\ind{}}{\ind{}'}.
\end{aligned}
\end{equation}

We thus see that, if we attempt to represent a Dicke-state density matrix without coherences, we must somehow get rid of the off-diagonal entries. Concretely, this is done by phase-averaging. We define the phase-averaged density matrix
\begin{equation}
\overline{\rho}(\varepsilon)^{\otimes N}=\int_{0}^{2\pi}\frac{d\phi}{2\pi}\,\rho(\varepsilon,\phi)^{\otimes N}.
\end{equation}

Since the phases average out for the off-diagonal terms
\begin{equation}
\int_{0}^{2\pi}\frac{d\phi}{2\pi}\,e^{i\phi\,(\ind{}-\ind{}')} = \delta_{\ind{},\ind{}'},
\end{equation}
we obtain a diagonal density matrix
\begin{equation}
\overline{\rho}(\varepsilon)^{\otimes N}=\sum_{\ind{}=0}^{N} \binom{N}{\ind{}}\, \varepsilon^\ind{}(1-\varepsilon)^{N-\ind{}}\,\outerproduct{\ind{}}{\ind{}}.
\end{equation}

Alternatively, a finite sum that accomplishes the same phase averaging is
\begin{equation}
\overline{\rho}(\varepsilon)^{\otimes N}=\frac{1}{N}\sum_{k=0}^{N-1}\rho\left(\varepsilon,\frac{2\pi k}{N}\right)^{\otimes N},
\end{equation}
which is the minimal number of terms that can cancel all phase terms for an arbitrary $\varepsilon\in[0,1]$.

Since the phase averaging is a technical detail, we will omit the notation indicating the phase averaging from now on, but assume it implicitly.

A separable, diagonal, symmetric state of $N$ such emitters can then be written as a convex combination of $\rho(\varepsilon)^{\otimes N}$. By diagonal, we mean that there is no single-emitter coherence since it is known that $\expval{\sigma_i}=0$ in the Dicke superradiance problem. Taking this into account, a separable state can thus be written as 
\begin{equation}
    \rho = \sum_{i=1}^M w_i\, \rho(\Ei{})^{\otimes N},
\end{equation}
with $w_i \geq 0$, $\sum_i w_i = 1$, and $\Ei{} \in [0,1]$.

These product density matrices are diagonal in the Dicke basis in the sense that they have matrix elements
\begin{equation}
    \bra{\Dicke{k}} \rho(\varepsilon)^{\otimes N} \Dickeket{} = \binom{N}{k} \varepsilon^k (1 - \varepsilon)^{N - k}.
\end{equation}

Thus, any separable, diagonal, symmetric state has Dicke populations of the form:
\begin{equation}
    p_\ind{} = \sum_{i=1}^M w_i \binom{N}{\ind{}} \Ei{}^\ind{} (1 - \Ei{})^{N - \ind{}}.
\end{equation}

This expression is a convex combination of Bernstein basis polynomials $b_{k,N}(\varepsilon) = \binom{N}{k} \varepsilon^k (1 - \varepsilon)^{N - k}$, which form a basis for the polynomials on the interval $[0,1]$. Instead of working with the Bernstein polynomials, it is more natural to work with the monomials. The Bernstein polynomials $\bm b(x)=(b_{0,N}(x),b_{1,N}(x),\dots,b_{N,N}(x))^T$ are related to the monomials $\bm v(x)=(1,x,\dots,x^N)^T$ with a linear transformation
\begin{equation}
    \bm v(x)=\bm B\bm b(x).
\end{equation}

This transformation matrix has matrix elements
\begin{equation}
    B_{\ind{},\ind'} = \binom{\ind{}'}{\ind{}}/\binom{N}{\ind{}}
\end{equation}
which defines an upper triangular matrix.

So, in terms of the moment vector $\bm m(t)=\bm B\bm p(t)$ the separability condition becomes 

\begin{equation}\label{eq:moment_equation}
    m_\ind{}(t) = \sum_i w_i(t) \Ei{}(t)^\ind{}.
\end{equation}

The sum $\sum_i w_i \Ei{}^k$ can now be considered as the $\ind{}$-th moment of the probability distribution $\sum_{i=1}^M w_i(t)\delta_{p,\Ei{}(t)}$. Asking whether $\Ei{}(t)\in[0,1]$ and convex $w_i(t)$ exist so that Eq.~\eqref{eq:moment_equation} is exactly equivalent to asking if a probability measure $f_t(p)$ supported on $[0,1]$ exists such that
\begin{equation}
    m_\ind{}(t)=\int_0^1 p^\ind{} f_t(p)\dd{p}\quad\text{for }k=0,1,2,\dots,N
\end{equation}
is fulfilled. This is the truncated moment problem on $[0,1]$ for each time $t$.

\section{The truncated Hausdorff moment problem}\label{app:hausdorff}

We adapt \cite{schmudgen2017moment} for this section. Given some probability distribution $\mu(x)$, all moments $m_{k}$ can be calculated from it. The moment problem solves the inverse problem. Given real numbers $m_0,m_1,\dots, m_N$, does there exist a positive measure $\mu$ on the interval $[0,1]$ such that
\begin{equation}
    m_\ind{}=\int_0^1 x^\ind{}\mu(x)\ \dd{x}\quad\text{for }\ind{}=0,1,2,\dots,N\text{ ?}
\end{equation}
The positive measure is called a probability distribution exactly when it is normalized, i.e., $m_0=1$.

To write the necessary and sufficient conditions, define the Hankel matrix of the moments
\begin{equation}
    H_{\ind{},\ind{}'}(\bm m)=m_{\ind{}+\ind{}'}
\end{equation}
with $\ind{},\ind{}'=0,1,\dots,\floor{N/2}$ and the localizing shifted Hankel matrix
\begin{equation}
    \overline{H}_{\ind{},\ind{}'}(\bm m)=m_{\ind{}+\ind{}'+1}-m_{\ind{}+\ind{}'+2}
\end{equation}
with $\ind{},\ind{}'=0,1,\dots,\floor{(N-1)/2}$.

\emph{Theorem:} A sequence $m_0,\dots,m_N$ corresponds to the moments of a probability distribution on the interval $[0,1]$ if and only if

\begin{itemize}
    \item The distribution is normalized $m_0=1$.
    \item The Hankel matrix $H(\bm m)$ and the shifted Hankel matrix $\overline{H}(\bm m)$ are positive semidefinite.
\end{itemize}

Since $H(\bm m)$ and $\overline{H}(\bm m)$ are symmetric Hankel matrices, they are positive semidefinite if and only if all their leading principal minors are nonnegative by Sylvester's theorem. That is to say, the determinants of the upper left $1\times 1$ $H_{1}(\bm m),\overline{H}_{1}(\bm m)$, the upper left $2\times 2$ block $H_{2}(\bm m),\overline{H}_{2}(\bm m)$ etc.~must all have nonnegative determinants. 

\section{Numerical reconstruction of supports and weights in the moment problem}\label{app:moment_reconstruction}

Given a moment vector $m_0,m_1,\dots,m_N$ we want to reconstruct the weights $w_i$ and $\Ei$ so that 
\begin{equation}\label{eq:moment_equation_constant}
    m_\ind{} = \sum_{i=1}^M w_i \Ei{}^\ind{}.
\end{equation}
where we know that $M\leq \ceil{(N+1)/2}$. In practice, this can be done without any nonlinear optimization using the following algorithm. The reconstruction procedure we use is standard in the classical theory of truncated moment problems \cite{schmudgen2017moment}, but to our knowledge, its application to the separability problem in quantum dynamics has not been systematically explored.

\begin{enumerate}
    \item Construct the Hankel matrix $H(\bm m)$ of the moment problem.
    \item Find the minimal $r$ so that the upper left $r\times r$ block $H_{r}(\bm m)$ has full rank via SVD and $H_{r+1}(\bm m)$ is singular up to $r=\ceil{(N+1)/2}$.
    \item Solve the linear system
    \begin{equation}
        H_{r}\bm{a} = -\bm{b}
    \end{equation}
    with $\bm{b} = (m_r, m_{r+1}, \dots, m_{2r-1})^T$ for $\bm a$.
    \item Construct a polynomial $p(x) = x^r + a_{r-1}x^{r-1} + \cdots + a_0$ and find its roots. Its real roots in the interval $[0,1]$ are the support points $x_i$.
    \item Construct the Vandermonde matrix $V_{ki} = x_i^k$ for $k = 0,\dots,r-1$ and solve the remaining linear system
    \begin{equation}
        V \bm{w} = (m_0, \dots, m_{r-1})^T
    \end{equation}
    for the weights $w_i$.
\end{enumerate}

Since the roots of the polynomial can be found as the eigenvalues of its companion matrix, the procedure reduces to linear algebra operations such as SVDs and diagonalization of matrices with dimension $\propto N$, leading to a computational complexity of $\mathcal{O}(N^3)$ for this algorithm.

\section{The moment problem for time evolution via linearization}\label{subsec:time}

To analyze whether separability is preserved under time evolution, we linearize the dynamics. Since studying the full exponential map is intractable, we ask whether a single small time-step already preserves the structure of a valid moment vector. If this property holds uniformly for all moment vectors for a small time step, it holds for all times by composition.

Let $\bm{m}(t)$ be a time-dependent moment vector evolving according to a matrix $\Momentmat{}$,
\begin{equation}
    \bm{m}(t) = \me^{t \Momentmat{}} \bm{m}(0).
\end{equation}

We now ask whether $\bm{m}(t)$ remains a valid moment vector for all $t \geq 0$ whenever $\bm{m}(0)$ is a valid moment vector. Showing that this full-time evolution maps proper moment vectors onto proper vectors is often tricky. Thus, rather than the full exponential, we linearize the time evolution 
\begin{equation}
    \bm{m} \mapsto \bm{m} + \deltat \Momentmat{} \bm{m}
\end{equation}
for small $\deltat>0$. If there exists some $\deltat$ for all $\bm m$ so that $\bm{m}(t)$ is a valid moment vector, then the full time evolution also only generates valid moment vectors for all times via
\begin{equation}\label{eq:semigroup}
    \bm m(t)=\lim_{n\to \infty}\left(\me^{\Momentmat{}t/n}\right)^n \bm m(0).
\end{equation}

A crucial point is that we do not need to check this for all $\bm m$. It is sufficient to check that all extremal points which are generated by delta distributions $\bm v(x)=(1,x,x^2,\dots,x^N)^T$ map onto valid moment distributions. This relates to the underlying density matrices in the following way: Checking all $\bm m(t)$ is equivalent to checking that the time evolution maps all separable density matrices onto separable density matrices. Checking only for $\bm v(x)$ amounts to showing that the time evolution maps all product density matrices onto separable states. But since separable states are only convex combinations of product density matrices, checking $\bm v(x)$ is sufficient. The first step in checking that $\Momentmat{}$ preserves moment vector validity is to show that
\begin{equation}
    \bm m(x,\deltat)=\bm v(x)+\deltat \Momentmat{}\bm v(x)
\end{equation}
remains a valid moment vector for all $x\in[0,1]$.

This first-order validity can be checked using the determinant-based conditions from the main text. For each $x$, we construct the Hankel and shifted Hankel matrices from $\bm m(x,\deltat)$, and verify that they are positive semidefinite to linear order in $\deltat$. By Sylvester’s theorem, this reduces to checking that all leading principal minors are nonnegative. To linearize the determinants, we use the fact that the derivative of a determinant of a square matrix $A$ can be written as
\begin{equation}\label{eq:linear_determinant}
    \det A(\deltat)=\det A(0)+\deltat \Tr(\operatorname{adj}(A(0))\cdot A'(0))+\mathcal O(\deltat^2),
\end{equation}
where $\operatorname{adj}$ indicates the adjugate of a matrix.

The problem now simplifies crucially by recognizing that both $H(\bm v(x))$ and $\overline H(\bm v(x))$ are rank-1 matrices: $H(\bm v(x))=\bm v(x)\bm v(x)^{\!T}$. For the shifted Hankel matrix define $\bm w(x)=\bigl((1-x),(1-x)x,(1-x)x^{2},\dots\bigr)$; then $\overline H(\bm v(x))=\bm w(x)\bm v(x)^{\!T}$. Consequently, \emph{at first order in $\deltat$}, the adjugate–trace term in Eq.~\eqref{eq:linear_determinant} vanishes for every principal minor except the $1{\times}1$ and $2{\times}2$ blocks. To complete the linear-order check, we therefore verify that
\[
\begin{aligned}
&{}\det H_1\!\bigl(\bm m(\deltat,x)\bigr),\;
\det\overline H_1\!\bigl(\bm m(\deltat,x)\bigr),\;\\
&{}\det H_2\!\bigl(\bm m(\deltat,x)\bigr),\;
\det\overline H_2\!\bigl(\bm m(\deltat,x)\bigr)
\end{aligned}
\]
remain non-negative. Their linearized forms read explicitly
\begin{widetext}
\begin{equation}\label{eq:linearized_ineq}
\begin{aligned}
    \det H_1(\bm m(x,\deltat)) &= 1 + \deltat\,\dot{m}_0(x,0)\geq 0, \\
\det H_2(\bm m(x,\deltat)) &= \deltat\left( \dot{m}_0(x,0)\,x^2 + \dot{m}_2(x,0) - 2x\,\dot{m}_1(x,0) \right)\geq 0, \\
\det \overline H_1(\bm m(x,\deltat)) &= x - x^2 + \deltat\left( \dot{m}_1(x,0) - \dot{m}_2(x,0) \right)\geq 0, \\
\det \overline H_2(\bm m(x,\deltat))&= \deltat x(1 - x)\left( \dot{m}_3(x,0) - \dot{m}_4(x,0)+ x^2\left( \dot{m}_1(x,0) - \dot{m}_2(x,0) \right)- 2x\left( \dot{m}_2(x,0) - \dot{m}_3(x,0) \right) \right)\geq 0\\
\end{aligned}
\end{equation}
\end{widetext}
and $m_0(x,\deltat)=1$.

It is sufficient to determine whether there exists some $\deltat>0$ so that these inequalities hold for all $x\in[0,1]$ to show that $H_1(\bm m(x,\deltat))$ and $H_2(\bm m(x,\deltat))$ remain positive for all times.

We now check the resulting inequalities for the specific moment time evolution matrix $\Momentmat{}$ associated with Dicke superradiance. We find the linearized time evolution of the moments as 
\begin{equation}\label{eq:linearized_moments}
    m_\ind{}(x,\deltat) = x^\ind{} + \deltat \ind{}(-N-1+\ind{}+(N-\ind{})x) x^\ind{}+\mathcal O(\deltat^2),
\end{equation}
so that
\begin{equation}
\dot m_\ind{}(x,0) = x^\ind{} \ind{}(-N-1+\ind{}+(N-\ind{})x)
\end{equation}
The first condition in Eq.~\eqref{eq:linearized_ineq} is trivially satisfied. The inequality for $\overline H_2(\bm m(x,\deltat))$ is equivalent to $x\geq 0$ and the inequality for $H_2(\bm m(x,\deltat))$ is equivalent to $x\leq 1$. Only the condition $\det \overline H_1(\bm m(x,\deltat))\geq 0$ constrains the time $\deltat$ for which these inequalities hold. It holds trivially for $x=1$ and $x=0$ and for $0<x<1$ it explicitly reads
\begin{equation}
\deltat \left(-2 N x^2+3Nx-N+4 x^2-3 x\right)\geq x-1
\end{equation}
which always holds for $\deltat<1/N$. We thus see that $\bm{I}+\deltat\Momentmat{}$ maps all valid moment vectors onto valid moment vectors for $\deltat\in[0,1/N]$. The limit to the full-time evolution in Eq.~\eqref{eq:semigroup} can be safely taken since the inequalities derived hold uniformly for all $x\in[0,1]$. To show that no entanglement is generated during Dicke time evolution, we must now show that the lowest nonzero time orders in all other $H_r(\bm m(x,\deltat))$ are also positive.

\section{Higher-Order Hankel Minors and Positivity}\label{app:all_minors}

For $r\ge 3$, the coefficient of the linear term in the time expansion of $\det H_r$ vanishes.  Hence we must show that the first \emph{non-zero} contribution---occurring at order $\deltat^{\,r(r-1)/2}$---is
strictly positive. For instance, completely independent of $N$, we obtain
\begin{equation}\label{eq:examples}
    \begin{aligned}
        H_3(\bm m(x,\deltat))&=16(1-x)^3x^6\deltat^3+\mathcal O(\deltat^4)\\
        H_4(\bm m(x,\deltat))&=768 (1-x)^6 x^{12}\deltat^{6}+\mathcal O(\deltat^7)
    \end{aligned}
\end{equation}
which are manifestly positive for $x\in(0,1)$.

\smallskip
\emph{Roadmap for this appendix.---}
\begin{enumerate}[nosep,label=(\roman*)]
  \item Expand $\det H_r$ in powers of $\deltat$ via multilinearity and identify the first non-zero order $\deltat^{\,r(r-1)/2}$.
  \item Factor that leading term into explicit powers of $x$ and $(1-x)$, leaving a coefficient $K_r$ independent of $x$.
  \item Prove $K_r>0$ for all $r$ and repeat the argument for the shifted Hankel minors $\overline H_r$, thereby completing the positivity proof.
\end{enumerate}
\smallskip

\emph{Moment generator.---} We compute the functional form of the time evolution generator for the moment vector
\begin{equation}
  \Momentmat = \begin{pmatrix}
    -\beta_0 &  \lambda_0 & 0 & \dots & 0 \\
    0 & -\beta_1 & \lambda_1 & \dots & 0 \\
    \vdots & \vdots & \ddots & \ddots & \vdots \\
    0 & 0 & \dots & -\beta_{N-1} & \lambda_{N-1} \\
    0 & 0 & \dots & 0 & -\beta_N
  \end{pmatrix},
\end{equation}
where all unspecified entries are equal to zero. The rates are explicitly given by $\lambda_{k}=k(N-k)$ and $\beta_k=k(N-k+1)$, see Eq.~\eqref{eq:linearized_moments}.

\emph{Moments and Hankel Matrix.---} For a parameter \(x \in (0,1)\), define the vector \(\bm v(x)\), the moment vector \(\bm m(x,\deltat)\), and its components \(m_k(x,\deltat)\) as:
\begin{equation}
\begin{aligned}
       \bm v(x) &= (1,x,x^2,\dots,x^N)^T,\\
       \bm m(x,\deltat)&=\me^{\deltat\Momentmat} \bm v(x),\\
       m_k(x,\deltat) &= (\bm m(x,\deltat))_k.
\end{aligned}
\end{equation}

The \emph{Hankel matrix} of size \(r\) constructed from these moments is:
\begin{equation}
   H_r(\bm m(x,\deltat)) = \bigl[m_{n+k}(x,\deltat)\bigr]_{k,n=0}^{r-1}.
\end{equation}

\subsection{Rank Structure of $H_{r}$ at fixed time order}\label{sec:rank}

Define the \emph{truncated moment vectors} as:
\begin{equation}
  \bm m^{[i]}(x) = \Momentmat^i \bm v(x), \quad \text{thus} \quad \bm m(x,\deltat) = \sum_{i=0}^\infty \frac{\deltat^n \bm m^{[i]}(x)}{i!}.
\end{equation}

Then, define the truncated Hankel submatrix of size $r$ and rank $n$
\begin{equation}
    H_r(\bm m^{[i]})=\left[m^{[i]}_{n+k}(x)\right]_{k,n=0}^{r-1}.
\end{equation}

We can write $m^{[i]}_k=x^k P_{i,k}(x)$ where $P_{i,k}(x)$ is a polynomial of degree $\deg P_{i,k}(x)\leq i$. This implies that $H_r(m^{[i]})$ has at most rank $i+1$ since you only have degrees of freedom $(x^k,x^{k+1},\dots,x^{k+i})$ to construct column vectors. 

\subsection{Determinant expansion}

Consider now the determinant

\begin{equation}
     \det H_r(\bm m(x,\deltat))=\det H_r\left(\sum_{i=0}^{\infty} \deltat^i \bm m^{[i]}(x)\right)
\end{equation}

By the multilinearity of the determinant, we can now expand each column of the determinant in orders of time
\begin{widetext}
\begin{equation}\label{eq:det-full-sum}
   \det H_r(\bm m(x,\deltat))=\sum_{\iind\ge0}
          \frac{\deltat^{\,i_0+i_1+\dots+i_{r-1}}}{i_0!\dots i_{r-1}!}
          \det H_r\bigl(\bm m^{[i_0]}(x),\dots,\bm m^{[i_{r-1}]}(x)\bigr)
\end{equation}
\end{widetext}
where in the last line $H_r\bigl(\bm m^{[i_0]}(x),\dots,\bm m^{[i_{r-1}]}(x)\bigr)$ is the matrix built with the $c$-th column given by the $c$-th column of $H_r\left(\bm m^{[n_i]}(x)\right)$. By a rank argument, we can immediately recognize that the lowest trivially nonvanishing order is any multi-index $\{\iind\}$ so that $\sum_{j}i_j=r(r-1)/2$. So

\begin{widetext}
\begin{equation}\label{eq:det-full-sum-order}
   \det H_r(\bm m(x,\deltat))=\deltat^{r(r-1)/2}\sum_{\substack{\iind\ge0 \\ \sum i_j=r(r-1)/2}}
          \frac{\det H_r\bigl(\bm m^{[i_0]}(x),\dots,\bm m^{[i_{r-1}]}(x)\bigr)}{i_0!\dots i_{r-1}!}+\mathcal O(\deltat^{r(r-1)/2+1})\\
\end{equation}
\end{widetext}

We now discuss the structure of this matrix determinant in detail and show a specific form.

\subsection{Extracting the $x^{r(r-1)}$ factor from $H_r$}

Our goal is to prove
\begin{equation}\label{eq:prove_me2}
\begin{aligned}
    \det H_r\!\bigl(\bm m(x,\deltat)\bigr)&=K_r\,\deltat^{r(r-1)/2}\,(1-x)^{r(r-1)/2}\,x^{r(r-1)}\\
   &+\mathcal O(\deltat^{r(r-1)/2+1}),\qquad K_r\in\mathbb R .
\end{aligned}
\end{equation}

\paragraph*{Column–row factorisation.}
Because the generator \(\Momentmat\) never lowers the degree in \(x\),
\begin{equation}
       m^{[i]}_{k}(x)=(-\beta_k)^{i}\,x^{k}+ \mathcal O\!\bigl(x^{k+1}\bigr).
\end{equation}
Hence the entry in row \(q\) and column \(c\) of \(H_r\) is proportional
to \(x^{q+c}\).
Factor \(x^{c}\) from column \(c\;(c=0,\dots,r-1)\) and \(x^{q}\) from
row \(q\;(q=0,\dots,r-1)\); the surviving core matrix is independent of
\(x\), while the total prefactor is
\[
    x^{\sum_{c=0}^{r-1}c}\,x^{\sum_{q=0}^{r-1}q}
   =x^{\,r(r-1)/2}\,x^{\,r(r-1)/2}=x^{\,r(r-1)}.
\]
Thus
\begin{equation}
   \det H_r\!\bigl(\bm m(x,\deltat)\bigr)=x^{\,r(r-1)}\,P_r(x,\deltat),
\end{equation}
where \(P_r(x,\deltat)\) is a polynomial in \(x\) of degree at most \(r(r-1)/2\).  The next subsection will factor out the remaining \((1-x)^{r(r-1)/2}\) and identify \(K_r\).

\onecolumngrid

\subsection{Extraction of $(1-x)^{r(r-1)/2}$ factor from $H_r$}

\vspace{4pt}
\noindent\emph{Steps in this subsection.---}%
\begin{enumerate}[leftmargin=*,label=(\arabic*)]
  \item \emph{Symmetrised minor.}  For each ordered index tuple
        $\mathbf i=(i_0,\dots,i_{r-1})$ with
        $\sum i_j=r(r-1)/2$, define
        \(
           \Delta_{\mathbf i}(r)=\sum_{\sigma\in S_r}
           \det H_r\bigl(\bm m^{[i_{\sigma(0)}]}(x),\dots\bigr)
        \)
        and write $\det H_r = \deltat^{r(r-1)/2}\!\!\sum_{\mathbf i}M(\mathbf i)^{-1}\Delta_{\mathbf i}$.
  \item \emph{$y$-expansion of entries.}  
        Factor $x^{k}$ from every matrix element and expand the remainder in $y:=1-x$:
        \[
          m^{[n]}_k(x)/x^k
          \;=\;\sum_{\ell=0}^{n} w(k,n,\ell)\,y^{\ell}.
        \]
  \item \emph{Degree bound.}  Show that
        $w(k,n,\ell)$ is a polynomial of degree $n+\ell$ in $k$.
  \item \emph{Vandermonde selection.}  
        Re-expand $\Delta_{\mathbf i}(r)$ in $y$ by powers of $\bm \li$ into $\Sigma^{\bm i}_{\bm\li}(r)$ and use antisymmetry of the Vandermonde determinant to prove that $\Sigma^{\bm i}_{\bm\li}(r)=0$ unless $\bm\li=\bm i$. Hence the only surviving term carries the prefactor $y^{\,\sum \ell_j}=y^{\,r(r-1)/2}$.
  \item \emph{Result.}  Combining this $y$-power with the previously
        extracted $x^{r(r-1)}$ yields
        \[
          \det H_r(\bm m(x,\deltat))
          =K_r\,\deltat^{r(r-1)/2}\,
          x^{r(r-1)}(1-x)^{r(r-1)/2}\!
          \bigl[1+\mathcal O(\deltat)\bigr],
        \]
        with the positive constant $K_r$ computed in the next subsection.
\end{enumerate}
\vspace{4pt}

\begin{table}[h]
\centering
\begin{tabular}{@{}lll@{}}
\toprule
\textbf{Symbol} & \textbf{Range} & \textbf{Role / stage in proof}\\
\midrule
$k$ & $0,\dots,r\!-\!1$ & Row index of the original Hankel matrix $H_r$ \\
$n$ & $0,\dots,r\!-\!1$ & Column index of the original Hankel matrix $H_r$ \\
$d$ & $0,\dots,r\!-\!1$ & Row index after polynomial expansion \\
$i_j$ & $i_j\in\mathbb{N}_0$, $\sum_j i_j=r(r-1)/2$ & Time‐expansion order for column $j$ (multi-index $\iind$) \\
$\ell_j$ & $0,\dots,i_j$ & $(1-x)$–expansion order for column $j$ (multi-index $\lind$) \\
\bottomrule
\end{tabular}
\caption{Index conventions used in the extraction of the $(1-x)^{r(r-1)/2}$ factor.}
\end{table}

Recalling the column–by–column expansion obtained in the previous subsection, the lowest non-vanishing time-order of the Hankel determinant can be written as
\[
   \det H_r(\bm m(x,\deltat))=\deltat^{r(r-1)/2}\sum_{\substack{\iind\ge0 \\ \sum i_j=r(r-1)/2}}\det H_r\bigl(\bm m^{[i_0]}(x),\dots,\bm m^{[i_{r-1}]}(x)\bigr)+\mathcal O(\deltat^{r(r-1)/2+1})
\]

We now pick a specific multi-index $\{\iind\}$. One valid choice in this order is $i_j=j$. However, any combination with $\sum_{j=0}^{r-1}i_j=r(r-1)/2$ contributes to this minimal time order. We define this determinant as
\begin{equation}
\begin{aligned}
\Delta_{\iind}(r)&=\sum_{\sigma\in S_r}\det H_r\bigl(\bm m^{[i_{\sigma(0)}]}(x),\dots,\bm m^{[i_{\sigma(r-1)}]}(x)\bigr)\\
 \det H_r(\bm m(x,\deltat))&=\deltat^{r(r-1)/2}\orderedsum{i}M(\iind)^{-1}\Delta_{\iind}(r).
\end{aligned}
\end{equation}
where $S_r$ is the permutation group on ${0,1,\dots,r-1}$. We had to divide out the factor
\begin{equation}
    M(\iind)=\prod_{\text{unique }s\in{}\{\iind\}}(\# s)!,
\end{equation}
where $\#s$ indicates how often the unique element appears in the index set to divide out the over-counting via the permutation group. For instance, the tuple $\{1,1,1\}$ is counted six times by the permutation group, but only appears once in the original sum.

We now write the truncated moments explicitly as
\begin{equation}
m_{k}^{[n]}(x)/x^k=\sum_{m=0}^{n}(-1)^{n-m}x^{m}\left(\prod_{s=k}^{\,k+m-1}\lambda_{s}\right)\sum_{\substack{s_{0},\dots,s_{m}\ge0 \\ s_{0}+\cdots+s_{m}=n-m}}\prod_{i=0}^{m}\beta_{k+i}^{\,s_{i}},
\end{equation}
where we already indicated the proper division of the column and row factors of $x$. We now expand the right-hand side in terms of $y=1-x$ and obtain
\begin{equation}
    \begin{aligned}
        m_{k}^{[n]}(x)/x^k&=\sum_{\li=0}^n w(k,n,\li)y^\li\\
        w(k,n,\li)&=\sum_{m=0}^{n}\binom{m}{\li}(-1)^{\,n-m+\li}\left(\prod_{s=k}^{\,k+m-1}\lambda_{s}\right)\sum_{\substack{\indexset{s}{m} \ge0 \\ s_{0}+\cdots+s_{m}=n-m}}\prod_{i=0}^{m}\beta_{k+i}^{\,s_{i}}
    \end{aligned}
\end{equation}

We now expand each of the truncated moment vectors $m^{[\sigma(i_{r-1})]}$ in $\Delta_{\iind}(r)$ in orders of $y=1-x$ and use the multilinearity of the determinant again where we shall denote the expansion orders by $\li_n$. It turns out to be convenient to permute the expansion order and moment truncation order at the same time, so we also group all of the multi-indices $\li_{\sigma(n)}$ with the indices $i_{\sigma(n)}$ so that we perform the outer summation over the  $\{\lind\}$
\[
    \Delta_{\iind}(r)=\sum_{\lind}\sum_{\sigma\in S_r}\det H_r\left(\dots,\sum_{k=0}^{r-1} x^{n+k} w(k+r-1,i_{\sigma(n)},\li_{\sigma(n)}),\dots\right).
\]
so that 
\begin{equation}\label{eq:delta_expansion}
    \Delta_{\iind}(r)=\sum_{\lind}\Sigma^{\iind}_{\lind}(r),
\end{equation}
where
\begin{equation}\label{eq:sigma_expansion}
    \Sigma^{\iind}_{\lind}=\sum_{\sigma\in S_r}\det\left[w(k+n,i_{\sigma(n)},\li_{\sigma(n)})\right]_{k,n=0}^{r-1}
\end{equation}

We will now go on to prove that $\Sigma^{\iind}_{\lind}$ when the two index sets coincide $\{\lind\}=\{\iind\}$. Using Eq.~\eqref{eq:delta_expansion} this already implies Eq.~\eqref{eq:prove_me2} since this leads to a factor of $y^{r(r-1)/2}=(1-x)^{r(r-1)/2}$.

To proceed with the analysis of Eq.~\eqref{eq:sigma_expansion}, we note that $w(k,n,\li)$ is a polynomial in $k$ of degree $n+\li$
\begin{equation}
    w(k,n,\li)=\sum_{i=0}^{n+\li}\alpha_{i}^{(n,\li)} k^i.
\end{equation}

\begin{proof}
To determine the degree of $w(k,n,\li)$ define the the forward difference operator
\[
    \Delta f(k)=f(k+1)-f(k)
\]
with $\Delta^{0}$ the identity and $\Delta^\li$ the $\li$-fold application. The $\li$-fold forward difference is just the alternating binomial difference, so that
\[
    w(k,n,\li)=(-1)^\li\Delta^\li\left[U_n(k)\right]
\]
with
\[
    U_n(k)=\sum_{m=0}^{n}(-1)^{\,n-m}\left(\prod_{s=k}^{\,k+m-1}\lambda_{s}\right)\sum_{\substack{s_{0},\dots,s_{m}\ge0 \\ s_{0}+\cdots+s_{m}=n-m}}\prod_{i=0}^{m}\beta_{k+i}^{\,s_{i}}
\]
So that $\deg w(k,n,\li)=\deg U_n(k)-\li$. Each of the $\lambda_k$ and $\beta_k$ increases the power by two, and we have $n$ applications at most in the sum. Thus 
\[
    \deg w(k,n,\li)\leq 2n-\li.
\]
This upper bound is not fulfilled since the prefactors of $k^{2n-\li},k^{2n-\li-1},\dots,k^{n+\li+1}$ vanish. This is because $k^{2n-\li}$ is constant in $m$, while $k^{2n-\li-1}$ is linear in $m$, and so on. Then, using the fact that
\[
 \sum_{m=\li}^{n}(-1)^{m-\li}\binom{m}{\li}\,m^{t}=0,
   \qquad\text{for all } 0\le t\le n-\li .
\]
we conclude that the first nonvanishing coefficient is $k^{n+\li}$ so that
\[
    \deg w(k,n,\li)=n+\li.
\]
\end{proof}

We use this polynomial expansion to rewrite $\Sigma^{\iind}_{\lind}$ as
\begin{equation}
    \Sigma^{\iind}_{\lind}=\sum_{\sigma\in S_r}\det\left[\sum_{i=0}^{i_{\sigma(n)}+\li_{\sigma(n)}}\alpha_{i}^{(i_{\sigma(n)},\li_{\sigma(n)})} (k+n)^i\right]_{k,n=0}^{r-1}.
\end{equation}

We then use the binomial theorem to write 
\[
    \Sigma^{\iind}_{\lind}=\sum_{\sigma\in S_r}\det\left[
        \sum_{d=0}^{i_{\sigma(n)} + \li_{\sigma(n)}} k^d\left[\sum_{i=d}^{i_{\sigma(n)} + \li_{\sigma(n)}}
        \binom{i}{d} n^{i-d}\alpha_{i}^{(i_{\sigma(n)},\li_{\sigma(n)})}\right]\right]_{k,n=0}^{r-1}.
\]

The upper limit of the sum $i_{\sigma(n)} + \li_{\sigma(n)}$ can be replaced by $r-1$ since the binomial factor cancels every term $d>i_{\sigma(n)} + \li_{\sigma(n)}$
\[
    \Sigma^{\iind}_{\lind}=\sum_{\sigma\in S_r}\det\left[
        \sum_{d=0}^{r-1} k^d\left[\sum_{i=d}^{i_{\sigma(n)} + \li_{\sigma(n)}}
        \binom{i}{d} n^{i-d}\alpha_{i}^{(i_{\sigma(n)},\li_{\sigma(n)})}\right]\right]_{k,n=0}^{r-1}.
\]

Since this is a determinant of a product of two matrices, it can be factorized, leading to
\begin{equation}
\Sigma^{\iind}_{\lind}=\det[k^d]_{k,d=0}^{r-1}\sum_{\sigma\in S_r}\det\left[\sum_{i=d}^{i_{\sigma(n)} + \li_{\sigma(n)}}\binom{i}{d} n^{i-d}\alpha_{i}^{(i_{\sigma(n)},\li_{\sigma(n)})}\right]_{d,n=0}^{r-1}
\end{equation}

We can now exchange the permutation of the indices with the permutation of $n$ via.
\[
    S^{\iind}_{\lind}=\sum_{\sigma\in S_r}\text{sgn}{\sigma}\det\left[\sum_{i=d}^{i_{n} + \li_{n}}\binom{i}{d} \sigma(n)^{i-d}\alpha_{i}^{(i_{n},\li_{n})}\right]_{d,n=0}^{r-1}
\]

We now expand this determinant by using the Leibniz formula
\[
    S^{\iind}_{\lind}=\sum_{\sigma \in S_r}\sgn(\sigma)\sum_{\tau \in S_r}\sgn(\tau)\prod_{n=0}^{r-1}\left[\sum_{i_n=\tau(n)}^{i_n+\li_n}\binom{i_n}{\tau(n)}\sigma(n)^{i_n-\tau(n)}\alpha_{i_n}^{i_n,\li_n}\right]
\]
and pull the summation over $\sigma$ into the product
\begin{equation*}
\begin{split}
        S^{\iind}_{\lind}&=\sum_{\pi\in S_r}\sum_{k_0=0}^{\,i_{\pi(0)}+\li_{\pi(0)}}\;\cdots\;\sum_{k_{r-1}=0}^{\,i_{\pi(r-1)}+\li_{\pi(r-1)}}\,\Bigl[\sgn(\pi)\,\prod_{d=0}^{r-1}\binom{k_d}{d}\,\alpha_{k_d}^{(i_{\pi(d)},\li_{\pi(d)})}\Bigr]\\
    &\times\underbrace{\sum_{\sigma\in S_r}\sgn(\sigma)\prod_{d}\sigma(\pi(d))^{\,k_d-d}}_{T_{\bm k,\sigma,\pi}}
\end{split}
\end{equation*}

We can now change variables to $\tau=\sigma\circ\pi$ so that
\[
    T_{\bm k,\pi,\sigma}= \sgn(\pi)^{-1}\sum_{\tau\in S_r}\sgn(\tau)\prod_{d=0}^{r-1}\tau(d)^{\,k_d-d},
\]

We now recognize that the summation over $\tau$ is a Leibniz expansion of the Vandermonde determinant
\[
    \sum_{\tau\in S_r}\sgn(\tau)
  \prod_{d=0}^{r-1}\tau(d)^{\,k_d-d}
= \det\!\bigl[n^{\,k_d-d}\bigr]_{d,n=0}^{r-1}.
\]
Using this in the expression for the symmetrized determinant, we obtain
\begin{equation}
S^{\iind}_{\lind}=\sum_{\pi\in S_r}\sum_{k_0=0}^{\,i_{\pi(0)}+\li_{\pi(0)}}\;\cdots\;\sum_{k_{r-1}=0}^{\,i_{\pi(r-1)}+\li_{\pi(r-1)}}\,\Bigl[\prod_{d=0}^{r-1}\binom{k_d}{d}\,\alpha_{k_d}^{(i_{\pi(d)},\li_{\pi(d)})}\Bigr]\det\!\bigl[n^{\,k_d-d}\bigr]_{d,n=0}^{r-1}.
\end{equation}
Note the exponents of the Vandermonde $e_d=k_d-d$ must collectively span $\{0,1,2,\dots,r-1\}$, see Eq.~\eqref{eq:equal_vandermonde}. Otherwise, the determinant is zero. Now, note that $\alpha_k^{n,\li}=0$ when $\li>n$. Thus $\li_n\leq i_n$ and index $k_d$ must run from $d\leq k_d\leq i_n+\li_n$. It follows that $0\leq e_d\leq i_n+\li_n$. So the only way to fulfill all of these bounds is to have $\{e_0,e_1,\dots,e_{r-1}\}=\{0,1,2,\dots,r-1\}$ and $i_n=\li_n=d$. Consequently, $k_d-2d$ is the only term that remains in the sum, and $i_n=\li_n$. Thus, we have that $\sum_n \li_n=r(r-1)/2$ and indeed, since this is the power of $y$ we have
\begin{equation}\label{eq:prove_men}
   \det H_r(\bm m(x,\deltat))=K_r(1-x)^{r(r-1)/2}x^{r(r-1)} \deltat^{r(r-1)/2}+\mathcal O\left(\deltat^{r(r-1)/2+1}\right)\quad K_r\in\mathbb{R},
\end{equation}
We can also write the determinant explicitly, which no longer depends on the specifics of the Dicke problem since $\alpha^{(d,d)}_{2d}=1$
\begin{equation}
  \Sigma^{\iind}_{\iind}
=\left(\prod_{m=0}^{r-1}m!\right)S^{\iind}_{\iind}
=\left(\prod_{m=0}^{r-1}m!\right)\sum_{\pi\in S_r}
\biggl[\prod_{d=0}^{r-1}\binom{2i_{\pi(d)}}{d}\biggr]\det\!\bigl[n^{\,2i_{\pi(d)}- d}\bigr]_{d,n=0}^{r-1}.
\end{equation}

\subsection{Positivity of the leading coefficient $K_r$}

Combining all of the prefactors that we set aside earlier to make the discussion more manageable, the contribution to $K_r$ from the index set $I$ is then defined as
\begin{equation}
K_{\iind}=M(\iind)^{-1}\left(\prod_{d=0}^{r-1}i_d!\right)^{-1}\left(\prod_{m=0}^{r-1}m!\right)\sum_{\pi\in S_r}
\biggl[\prod_{d=0}^{r-1}\binom{2i_{\pi(d)}}{d}\biggr]\det\!\bigl[n^{2i_{\pi(d)} - d}\bigr]_{d,n=0}^{r-1}
\end{equation}
where we inserted the multiplicities $M(\iind)^{-1}$ that were divided out earlier and the factorials from the time evolution $\left(\prod_{d=0}^{r-1}i_d!\right)^{-1}$ to obtain the sum
\begin{equation}
    K_r=\orderedsum{i}K_{\iind}=2^{r(r-1)/2}\prod_{k=1}^{r-1}k!>0
\end{equation}

\begin{proof}
The first step is to move to the unordered sum, accounting for the multiplicities appropriately
\[
K_r=\frac{\prod_{m=0}^{r-1}m!}{r!}\sum_{\pi\in S_r}
\sum_{\substack{\iind\ge0\\\sum i_j=r(r-1)/2}}
\left[\prod_{d=0}^{r-1}\binom{2i_{\pi(d)}}{d}/i_{\pi (d)}!\right]\det\!\left[n^{2i_{\pi(d)} - d}\right]_{d,n=0}^{r-1}
\]
since every index tuple appears $r!M(\iind)^{-1}$ more often in the unordered sum than in the ordered sum (since we sum over all index set permutations, these contributions are all equal). We can then use the multilinearity of the determinant to pull the prefactors and sum over $\{\iind\}$ into the determinant
\[
K_r=\frac{\prod_{m=0}^{r-1}m!}{r!}\sum_{\pi\in S_r}
\sum_{\substack{\iind\ge0\\\sum i_j=r(r-1)/2}}
\det\!\left[\binom{2i_{\pi(d)}}{d}\frac{n^{2i_{\pi(d)} - d}}{i_{\pi(d)}!}\right]_{d,n=0}^{r-1}
\]

We now drop the $\pi$-label in the sum and execute the outer permutation sum 
\[
K_r=\left(\prod_{m=0}^{r-1}m!\right)
\sum_{\substack{\iind\ge0\\\sum i_j=r(r-1)/2}}
\det\!\left[\binom{2i_d}{d}\frac{n^{2i_{d} - d}}{i_d!}\right]_{d,n=0}^{r-1}
\]

We would now like to pull the index sum into the determinant. But the constraint $\sum i_j=r(r-1)/2$ prevents us from doing this. To get around this, define an ordinary generating function
\[
    G(\lambda)=\left(\prod_{m=0}^{r-1}m!\right)
\sum_{\iind\ge0}\det\!\left[\binom{2i_d}{d}\frac{n^{2i_{d} - d}}{i_d!}\right]_{d,n=0}^{r-1}\lambda^{\sum i_j}
\]
so that $K_r$ can be extracted as the $r(r-1)/2$-order series coefficient. We can now use the multilinearity of the determinant 
\[
    G(\lambda)=\left(\prod_{m=0}^{r-1}m!\right)\det\!\left[\sum_{k=0}^\infty \binom{2k}{d}\frac{n^{2k - d}}{k!}\lambda^k \right]_{d,n=0}^{r-1}
    =\left(\prod_{m=0}^{r-1}m!\right)\det\!\left[\frac{1}{d!}\dv[d]{n}\me^{n^2\lambda}\right]_{d,n=0}^{r-1}
\]

We now recognize that the argument of the determinant has the form
\[
    \frac{1}{d!}\dv[d]{n}\me^{n^2\lambda}=\me^{n^2\lambda}\left[\frac{(2\lambda)^d}{d!}n^d+R(n)\right]
\]
where $R(n)$ is a lower degree polynomial in $n$. The lower-degree polynomial $R(m)$ does not contribute to the determinant by the standard Vandermonde argument. Then, by the usual Vandermonde determinant formula for polynomial entries (Eq.~\eqref{eq:polynomial_vandermonde}), we obtain
\[
    G(\lambda)=\left(\prod_{m=0}^{r-1}m!\right)\det\!\left[\me^{n^2\lambda}\left[\frac{(2\lambda)^d}{d!}n^d+R(n)\right]\right]_{d,n=0}^{r-1}=\left(\prod_{m=0}^{r-1}m!\right)(2\lambda)^{r(r-1)/2}\exp\left(\lambda\frac{(2r-1)(r-1)r}{6}\right)
\]

We can now read off, as the series coefficient $[\lambda^{r(r-1)/2}]$ of $G(\lambda)$, the final result
\[
    K_r=2^{r(r-1)/2}\left(\prod_{m=0}^{r-1}m!\right)\ge0
\]
as claimed.
\end{proof}

\twocolumngrid

\subsection{Shifted Hankel minors $\overline H_r$}\label{sec:shifted}
Recalling the forward difference operator
\begin{equation}
    (\Delta \bm m(x,\deltat))_k \;:=\; m_{k+2}(x,\deltat)-m_{k+1}(x,\deltat),
\end{equation}
with $k=0,1,\dots,N-2,$ and the shifted Hankel matrix
\begin{equation}
   \overline H_r(\bm m(x,\deltat))=\bigl[(\Delta \bm m(x,\deltat)_{i+j}\bigr]_{i,j=0}^{r-1}.    
\end{equation}

In principle, the entire argumentation for the Hankel matrix also works for the shifted Hankel matrix since the shift operator commutes with the time evolution
 \begin{equation}
\begin{aligned}
\Delta \bm m(x,\deltat)&=\Delta\me^{\Momentmat\deltat}\bm v=\me^{\Momentmat\deltat}\Delta\bm v(x)\\
&=x(1-x)\me^{\Momentmat\deltat}\bm v(x)=x(1-x)\bm m(x,\deltat).
\end{aligned}
\end{equation}

So we obtain an additional prefactor of $1-x$. In total, we then obtain
\begin{equation}
\begin{aligned}
\det \overline H_r(\bm m(x,\deltat))&=K_rx^{r^2}(1-x)^{r(r+1)/2}\deltat^{r(r-1)/2}\\
&+\mathcal O\left(\deltat^{r(r-1)/2+1}\right)    
\end{aligned}
\end{equation}
with the same $K_r$ as before --- the shift contributes exactly a factor of $x(1-x)$, leading to $x^{r(r-1)}(1-x)^{r(r-1)/2}\rightarrow x^{r^2}(1-x)^{r(r+1)/2}$.

\subsection{The points $x=1$ and $x=0$}

At $x=0$ nothing happens. All Hankel matrices are manifestly zero and will remain zero. For $x=1$ the story is different. In principle, one also has to ensure that the first nonvanishing matrix elements at the point $x=1$ are nonnegative. With the same strategy as before we can recognize that the first nonvanishing order in $x=1$ is

\begin{equation}
\begin{aligned}
    \det \overline H_r(\bm m(x,\deltat))&=K_rx^{r^2}(1-x)^{r(r+1)/2}\deltat^{r(r-1)/2}\\
    &+\mathcal O\left(\deltat^{r(r-1)/2+1}\right)
\end{aligned}
\end{equation}

which is also manifestly positive. As such, we have two series, one that makes the exponential uniformly convergent on the interior points and one that makes it uniformly convergent at $x=1$. Together, this ensures uniform convergence of the positivity of the Hankel matrices.

\subsection{Combining all orders: completion of the proof}\label{sec:cone}

We have shown explicitly in the linear order earlier that the defining inequalities for moment vectors on the interval $[0,1]$ are fulfilled uniformly in $[0,1]$. Using the exact expansions for both the ordinary and shifted Hankel minors we derive
\begin{equation}
\begin{aligned}
\det \overline H_r(\bm m(x,\deltat))&=(1-x)^{r(r+1)/2}x^{r(r-1)}K_r \deltat^{r(r-1)/2}\\
&+\mathcal O\left(\deltat^{r(r-1)/2+1}\right)\\
   \det H_r(\bm m(x,\deltat))&=(1-x)^{r(r-1)/2}x^{r(r-1)}K_r \deltat^{r(r-1)/2}\\
   &+\mathcal O\left(\deltat^{r(r-1)/2+1}\right),
\end{aligned}
\end{equation}
with $K_r>0$ (Compare with the examples in Eq.~\eqref{eq:examples}). Therefore, there exists some $\delta>0$ so that $\me^{\Momentmat \delta}\bm v(x)$ remains a valid moment vector. Since any moment vector is a convex combination of such $v(x)$ we have thus proven that $\me^{\Momentmat \delta}$ maps valid moment vectors onto valid moment vectors. By composing small time ,steps we obtain conservation for all times. The uniformity of the bound in $x$ implies the convergence.

\section{Quantifying Entanglement via Moment Matrices}\label{subsec:entanglement_measure}

Beyond the analytical proof outlined above, the moment problem also yields a natural quantitative measure of entanglement, which corresponds to how much the conditions for the moment vector to be defined by a probability distribution on $[0,1]$ are violated. Define the negativity of a matrix as minus the sum over the absolute values of the negative eigenvalues
\begin{equation}
    \mathcal N(M)=-\sum_{\lambda \in \mathcal{S}(M),\lambda<0}\lambda,
\end{equation}
where we denoted the spectrum of $M$ as $\mathcal{S}(M)$. Then we define an entanglement measure of a population vector $\bm p$ as
\begin{equation}
    \mathcal N(\bm p)=\mathcal N(H(\bm B^{-1}\bm p))+\mathcal N(\overline H(\bm B^{-1}\bm p)).
\end{equation}

This measure is guaranteed to be precisely zero when the state is separable and nonzero for entangled states.

\begin{figure}[htbp]
    \centering
    \includegraphics[width=0.99\linewidth]{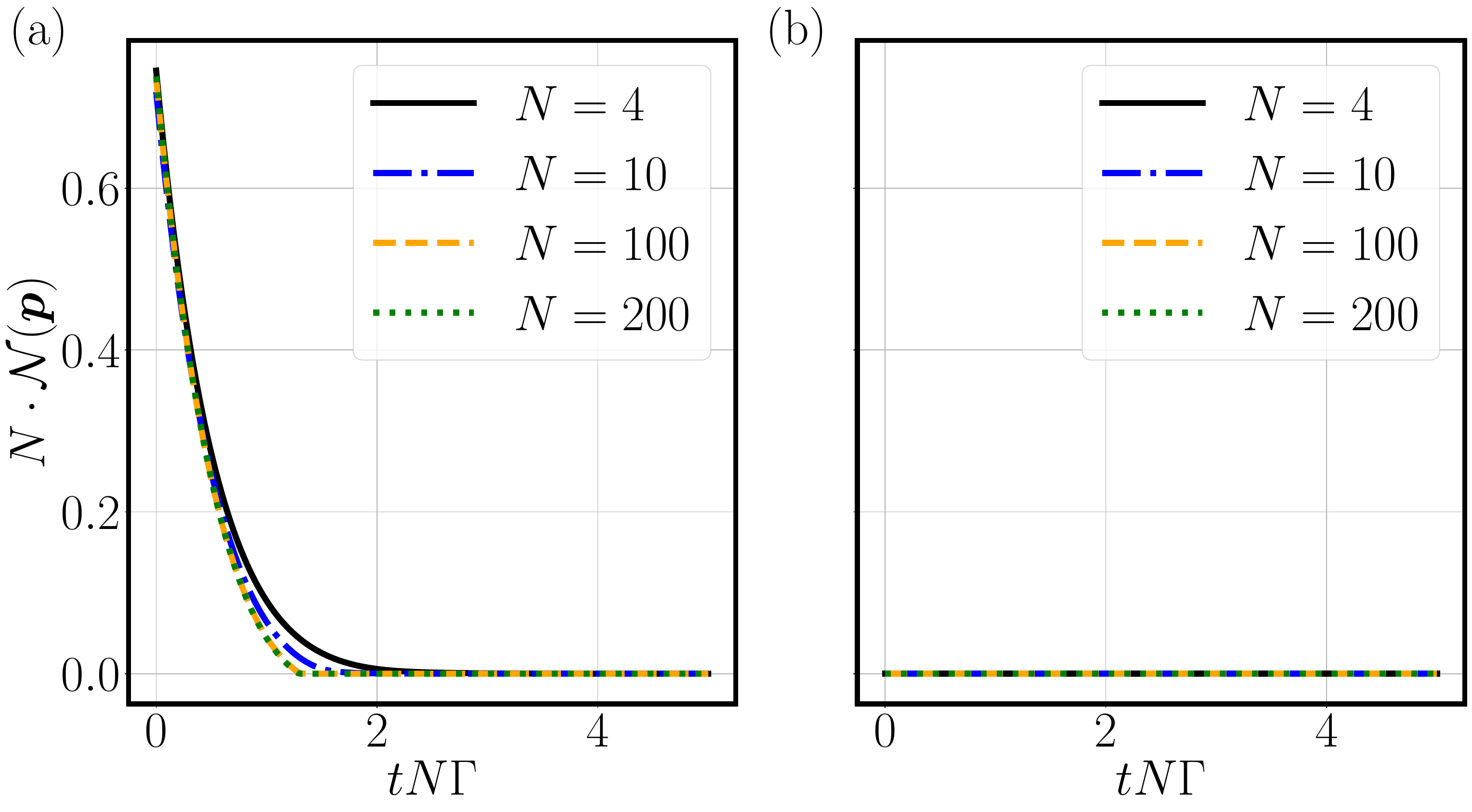}
    \caption{Here we show the time evolution of the Hankel matrix based entanglement measure $\mathcal N(\bm p)$ starting from the Dicke state $\ket{\Dicke{\floor{N/2}}}$ in (a) and for the fully inverted ensemble in (b). The time axis and negativity measure have been rescaled by $N$, to show universality for $N\to\infty$ in (a). The negativity for an initially excited state in (b) is manifestly zero for all times.}
    \label{fig:Hankel}
\end{figure}

We study this entanglement measure in Fig.~\ref{fig:Hankel} starting with half-inversion $\ket{\Dicke{\floor{N/2}}}$ in (a) and full inversion in (b). We can see that this entanglement measure remains manifestly zero when starting with the fully inverted state, confirming the earlier proof numerically. In (a), we can see that the entanglement measure decreases exponentially over time. Rescaling the time axis by $N$ and rescaling $\mathcal N(\bm p)$ by $N$ shows that this entanglement measure does seem to exhibit a universal curve shape as $N\to\infty$. This suggests an entanglement dilution rate that scales with $N$, and the entanglement of the initial Dicke state scales with $N$. 

\section{Two-Particle Entanglement and Monotonicity}\label{app::negativity}

In this appendix, we will discuss bipartite entanglement, in particular the entanglement between two representative spins of the permutationally invariant state, and evaluate the negativity of the resulting density matrix. This will give us good intuition about the entanglement while the system evolves according to the rate equations given in the main text.

\subsection{Two-emitter reduced state}\label{subsec:two_emitter_reduced}

Due to permutation symmetry, the reduced state of any two atoms is the same. For a given Dicke state $\Dickeket{}$, the two-emitter reduced density matrix $\rho_{12}^{(\ind{})}$ can be written in the basis $\{\ket{00}, \ket{01}, \ket{10}, \ket{11}\}$ as
\begin{equation}
    \rho_{12}^{(m)} =
    \begin{pmatrix}
        a_\ind{} & 0 & 0 & 0 \\
        0 & b_\ind{} & b_\ind{} & 0 \\
        0 & b_\ind{} & b_\ind{} & 0 \\
        0 & 0 & 0 & d_\ind{}
    \end{pmatrix},
\end{equation}
with coefficients
\begin{equation}
    \begin{aligned}
    a_\ind{} &= \expval{N_g^{(1)}N_g^{(2)}}=\frac{(N - \ind{})(N - \ind{} - 1)}{N(N - 1)}, \\
    b_\ind{} &= \expval{N_g^{(1)}N_e^{(2)}}=\frac{\ind{}(N - \ind{})}{N(N - 1)}, \\
    d_\ind{} &= \expval{N_e^{(1)}N_e^{(2)}}=\frac{\ind{}(\ind{} - 1)}{N(N - 1)}.
    \end{aligned}
\end{equation}
where we used the fact that emitter $1$ and emitter $2$ are representative for all emitters with $N_e^{i}=\sigma_i^\dagger\sigma^{\phantom\dagger}_i$ and $N_g^{i}=\sigma^{\phantom\dagger}_i\sigma_i^\dagger$.

The reduced state for a statistical mixture $\rho(t) = \sum_m p_m(t) \ket{S,m}\bra{S,m}$ is then given by
\begin{equation}\label{eq:rho12}
    \rho_{12}(t) = \sum_\ind{} p_\ind{}(t) \, \rho_{12}^{(\ind{})}=
    \begin{pmatrix}
        A & 0 & 0 & 0 \\
        0 & B & B & 0 \\
        0 & B & B & 0 \\
        0 & 0 & 0 & D
    \end{pmatrix},
\end{equation}
where $A = \sum_\ind{} p_\ind{} a_\ind{}$, $B = \sum_\ind{} p_\ind{} b_\ind{}$, and $D = \sum_\ind{} p_\ind{} d_\ind{}$. This matrix retains the same block-diagonal structure and depends only on the populations $p_\ind{}(t)$.

\subsection{Negativity}\label{subsec:negativity}

To quantify two-emitter entanglement, we consider the negativity $\mathcal N_2(t)$ of the reduced state $\rho_{12}(t)$, defined as
\begin{equation}
    \mathcal N_2(t) = \frac{\norm{\rho_{12}^{T_2}(t)}_1 - 1}{2},
\end{equation}
where $\rho_{12}^{T_2}(t)$ denotes the partial transpose with respect to the second emitter, and $\norm{\cdot}_1$ is the trace norm.

For a state of the form Eq.~\eqref{eq:rho12}, the partial transpose reads
\begin{equation}
    \rho_{12}^{T_2}(t) =
    \begin{pmatrix}
        A & 0 & 0 & B \\
        0 & B & 0 & 0 \\
        0 & 0 & B & 0 \\
        B & 0 & 0 & D
    \end{pmatrix}.
\end{equation}
The eigenvalues of $\rho_{12}^{T_2}(t)$ are $B$, $B$, and
\begin{equation}
    \lambda_\pm = \frac{A + D}{2} \pm \frac{1}{2}\sqrt{(A - D)^2 + 4B^2}.
\end{equation}
The only possibly negative eigenvalue is $\lambda_-$, so the negativity simplifies to
\begin{equation}\label{eq:twospin_neg}
\begin{split}
    \mathcal N_2(t) &= \max\left\{ 0, -\lambda_- \right\} \\
    &= \max\left\{ 0, \frac{\sqrt{(A - D)^2 + 4B^2} - (A + D)}{2} \right\}.
    \end{split}
\end{equation}

\subsection{Monotonicity}\label{subsec:Monotonicity}

Consider the possibly negative eigenvalue
\begin{equation}
    -\lambda_-=\frac{\sqrt{(A - D)^2 + 4B^2} - (A + D)}{2}
\end{equation}
then requiring that $\lambda_-< 0$ is equivalent to $A(t)D(t)-B(t)^2<0$. Thus, it is sufficient to study the sign of
\begin{equation}
    \Delta(t)=A(t)D(t)-B(t)^2
\end{equation}
to decide whether a state has finite negativity or not. Indeed, using the rate equations we show later that $\dot\Delta(t)\leq 0$, implying that the negativity decreases monotonically over time.

\subsection{Monotonicity of $\Delta$}\label{app:mono}

Using the Dicke rate equations, we compute
\begin{align*}
    \dot{A}(t) &= \sum_{\ind{}=0}^N h_\ind{} \, p_\ind{}(t) \, [a_{\ind{}-1} - a_\ind{}], \\
    \dot{B}(t) &= \sum_{\ind{}=0}^N h_\ind{} \, p_\ind{}(t) \, [b_{\ind{}-1} - b_\ind{}], \\
    \dot{D}(t) &= \sum_{\ind{}=0}^N h_\ind{} \, p_\ind{}(t) \, [d_{\ind{}-1} - d_\ind{}].
\end{align*}
Note that the $\ind{}=0$ term vanishes, so we formally keep it although there is no physical meaning for $a_{-1},b_{-1},d_{-1}$ since $h_0=0$.

From the explicit expressions
\begin{align*}
    a_{\ind{}-1} - a_\ind{} &= \frac{2(N - \ind{})}{N(N-1)}, \\
    b_{\ind{}-1} - b_\ind{} &= \frac{2\ind{} - N - 1}{N(N-1)}, \\
    d_{\ind{}-1} - d_\ind{} &= -\frac{2(\ind{} - 1)}{N(N-1)},
\end{align*}
we obtain
\begin{align*}
    \dot{A}(t) &= \frac{2}{N(N-1)} \sum_\ind{} h_\ind{} p_\ind{} (N -\ind{}), \\
    \dot{B}(t) &= \frac{1}{N(N-1)} \sum_\ind{} h_\ind{} p_\ind{} (2\ind{} - N - 1), \\
    \dot{D}(t) &= -\frac{2}{N(N-1)} \sum_\ind{} h_\ind{} p_\ind{} (\ind{} - 1).
\end{align*}

The equation of motion for $\Delta(t)$ then becomes
\begin{equation}
    \dv{\Delta}{t} = 2B \dot{B} - A \dot{D} - D \dot{A},
\end{equation}
where we can use the explicit representations for $A(t),B(t)$ and $D(t)$ to find

\begin{equation}
\begin{split}    
    &{}N^2(N-1)^2\dv{\Delta}{t}\\
    &=\sum_{\ind{},\ind{}'=0}^N
    p_{\ind{}'} p_\ind{}[2 \ind{} (N-1) (\ind{}-N+1) \\
    &{}\hphantom{\sum_{\ind{},\ind{}'=0}^N}\times(\ind{}' (N-1)+(1-\ind{})N)]\\
    &\leq 2 N (N-1)\sum_{\ind{},\ind{}'=0}^N
    p_{\ind{}'}p_\ind{}\left[(\ind{}' (N-1)+(1-\ind{})N)\right]\\
    &=2 N (N-1)\sum_{\ind{}=0}^N 
    p_\ind{}\left[(\ind{} (N-1)+(1-\ind{})N)\right]\\
    &=2 N (N-1)\sum_{\ind{}=0}^Np_\ind{}(\ind{}-N)\leq 0.
    \end{split}
\end{equation}

\subsection{Numerical Illustration}\label{subsec:numerica_illustration}

\begin{figure}[htbp]
\centering
    \includegraphics[width=0.99\linewidth]{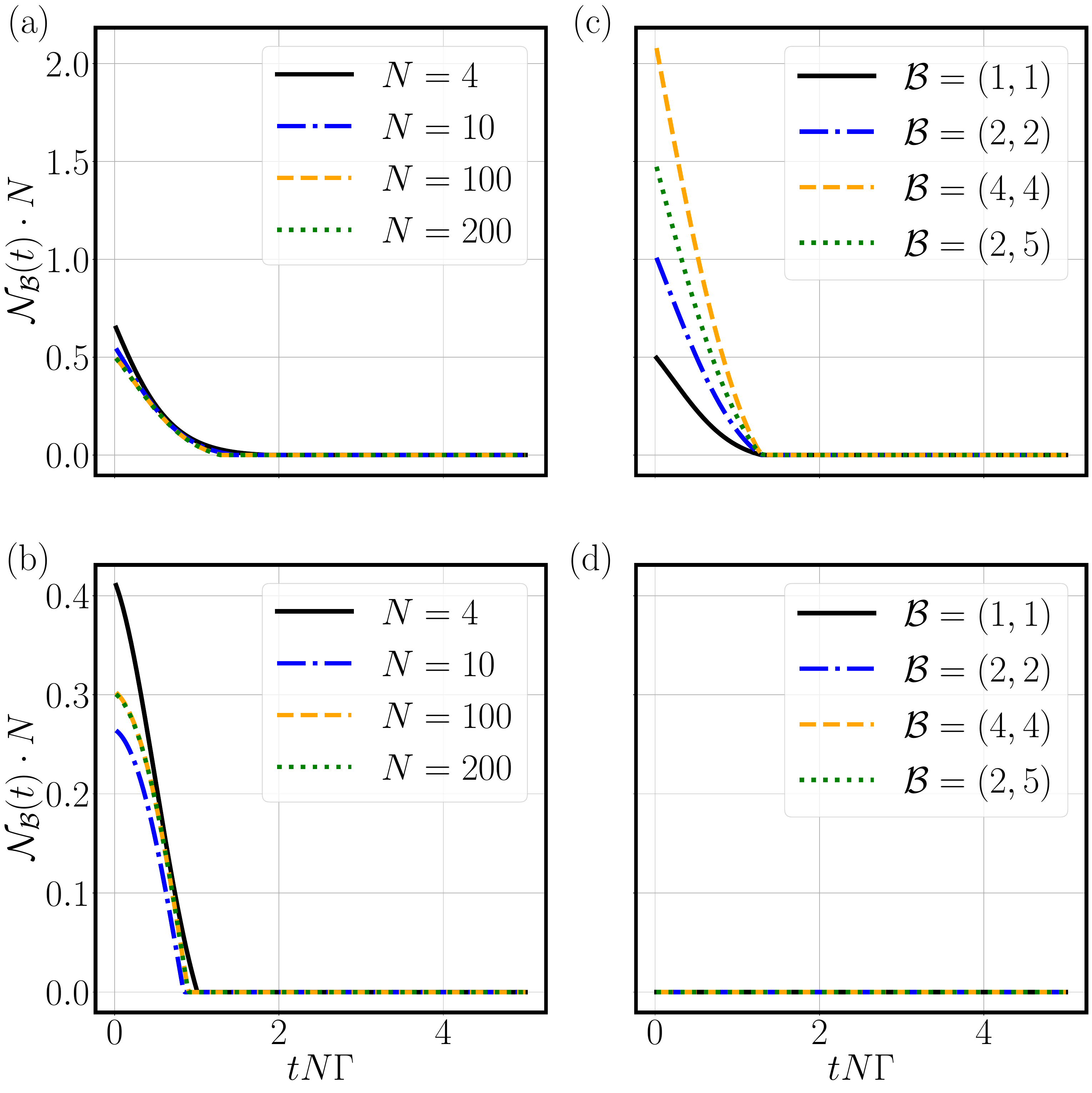}
    \caption{We show here plots of the negativity in different scenarios. In (a), we start from the Dicke state with a half-inverted ensemble and show the two-spin negativity defined in Eq.~\eqref{eq:twospin_neg} for different particle numbers $N$ and rescale the time and negativity by $N$ to show universality. In (b), we show the same quantity, but starting with the Dicke state where $3/4$ of the particles are inverted. In (c), we plot the negativity for $N=100$ for the density matrix for an effective density matrix partitioned as in $\mathcal B=(\mathcal B_1,\mathcal B_2)$, where $\mathcal B_1$ and $\mathcal B_1$ are the particle numbers in the respective subsystems and $\mathcal B_1+\mathcal B_2$ is the total particle number of the reduced density matrix for which the negativity is calculated. In (d), we show for $N=100$ that starting from full inversion, several bipartitions remain zero for all times. Details on how to calculate the reduced density matrices are detailed in Eq.~\eqref{eq:particle loss}.}
    \label{fig:negativity}
\end{figure}

To complement the analytic result, we conducted several numerical studies of negativity entanglement measures in Fig.~\ref{fig:negativity}. In Fig.~\ref{fig:negativity}(a,b), we can see that the two-spin negativity indeed decreases monotonically over time and becomes zero. More so, rescaling the negativity and time by the particle number shows that for large $N$ the negativity curves show universal behaviour that only depends on the initial state. Consequently, there is a characteristic time that scales linearly with the number of emitters after which no bipartite entanglement is left in the system. Strong evidence for this statement can be seen in Fig.~\ref{fig:negativity}(c), where we see that even for more complex bipartitions, the negativity decreases monotonously and vanishes at the same time for all bipartitions, even though the curve no longer has a universal shape. In (d), we confirm for some exemplary bipartitions, starting from the initially fully excited state, that no bipartite entanglement is ever generated via the Dicke superradiance time evolution.

\subsection{Calculation of reduced N spin density matrices for Dicke superradiance}\label{app:reduced}

Here we want to give a short outline on how the $m$-spin density matrix can be calculated for $N>m$ spins which are in a mixture of Dicke states. We take inspiration from Refs.~\cite{chase2008,zhang2018} and interpret this tracing procedure as a quantum jump that occurs due to particle loss from the system. The exact matrix elements of this operator in the Dicke basis are given in these references.

We state the tracing procedure in terms of states $\ket{N,S,m}$ for particle number $N$, spin sector $S$, and magnetic quantum number $m$. The nonzero matrix elements of the particle loss superoperator $\mathcal L_p$ starting from the fully symmetric subspace (which we always have) are precisely
\begin{equation}
    \label{eq:particle loss}
    \begin{split}
          \mathcal L_p\ket{N,S,m}\bra{N,S,m}&=
    \lambda_{\downarrow,N,S,M}\outerproduct{\downarrow}{\downarrow}\\
    &+\lambda_{\uparrow,N,S,M}\outerproduct{\uparrow}{\uparrow} 
    \end{split}
\end{equation}
with 
\[\begin{aligned}
\lambda_{\downarrow,N,S,M}&=\frac{\left(N/2 + S + 1\right)(S - M)}{N (2S + 1)},\\
\lambda_{\uparrow,N,S,M}&=\frac{\left(N/2 + S + 1\right)(S + M)}{N (2S + 1)},\\
\ket{\downarrow}&=\ket{N-1,S-1/2,m-1/2},\\
\ket{\uparrow}&=\ket{N-1,S-1/2,m+1/2}.
\end{aligned}\] 

The $m$-particle density matrix can be obtained by applying $\mathcal L_p^{N-m}$ to the initial density matrix and then constructing the density matrix in the product basis by symmetrizing the state with the appropriate number of excited emitters. This reproduces the formula Eq.~\eqref{eq:rho12} for the reduced two-spin density matrix from the main text.

\section{Review of determinant algebra}\label{app:det}
Throughout, $\sigma\in S_r$ denotes a permutation of $\{0,\dots,r-1\}$.

\paragraph{A.1  Leibniz formula}
\[
  \det A \;=\;\sum_{\sigma\in S_r}\!\sgn(\sigma)
  \prod_{d=0}^{r-1} A_{d,\sigma(d)} .
\]

\paragraph{A.2  Multilinearity.}
For any scalars $c_d$ and column vectors $v_d$,
\[
  \det[c_0v_0,\dots,c_{r-1}v_{r-1}]
  \;=\;\Bigl(\prod_{d=0}^{r-1}c_d\Bigr)\det[v_0,\dots,v_{r-1}].
\]

\paragraph{A.3  Vandermonde determinant.}
For any distinct numbers \(x_0,\dots,x_{r-1}\),
\[
  \det[x_n^{\,d}]_{d,n=0}^{r-1}
  \;=\;\prod_{0\le d < n \le r-1} (x_n-x_d).
\]

\emph{Corollary (equally spaced nodes).}
With \(x_n=n\) one obtains the closed form
\begin{equation}\label{eq:equal_vandermonde}
      \det[n^{\,d}]_{d,n=0}^{r-1}
  \;=\;\prod_{0\le d<n\le r-1}(n-d)
  \;=\;\prod_{k=1}^{r-1} k!
\end{equation}

\paragraph{A.4  Triangular–Vandermonde factorisation.}
If each row is a degree-\(\le d\) polynomial,
\(A_{d,n}=P_d(n)=\sum_{k=0}^{d}a_{d,k}n^{k}\),
then
\[
  A=CV,\qquad
  C_{d,k}=a_{d,k}\ (\text{lower-triangular}),\quad
  V_{k,n}=n^{k},
\]
so
\begin{equation}\label{eq:polynomial_vandermonde}
  \det A=\Bigl(\prod_{d=0}^{r-1}a_{d,d}\Bigr)
          \prod_{0\le d<n\le r-1}(n-d).    
\end{equation}

Notably, only the prefactors of the highest-degree monomial contribute, and the result vanishes when the degree of the polynomial is smaller than $d$.

\clearpage
\bibliographystyle{apsrev4-1-custom}
\bibliography{references}

\end{document}